# Measuring Corruption from Text Data[*]

Arieda Muço

23 November 2025

**Abstract**

Using Brazilian municipal audit reports, I construct an automated corruption index that combines a dictionary of audit irregularities with principal component analysis. The index validates strongly against independent human coders, explaining 71–73 % of the variation in hand-coded corruption counts in samples where coders themselves exhibit high agreement, and the results are robust within these validation samples. The index behaves as theory predicts, correlating with municipal characteristics that prior research links to corruption. Supervised learning alternatives yield nearly identical municipal rankings ($R^2 = 0.98$), confirming that the dictionary approach captures the same underlying construct. The method scales to the full audit corpus and offers advantages over both manual coding and Large Language Models (LLMs) in transparency, cost, and long-run replicability.

[*] Central European University (e-mail: arieda.muco@gmail.com). I thank Olle Folke, Marc Kaufman, and David Strömberg for input on this paper; Claudio Ferraz, Frederico Finan, Francisco Garfias, and Jeffrey Timmons for sharing their data; and Ricardo Lima for research assistance. This work was initially a chapter of my PhD dissertation and incorporated in an early working-paper version of *"Learn from thy Neighbor: Do Voters Associate Corruption with Political Parties?"* (2017), which later evolved into the article *The Politician, the Party, the President: How Do Political Scandals Propagate Across the Party Network?* (2025), that uses this corruption measure to examine heterogeneity patterns. This paper revises and extends that earlier methodological idea. Financial support from Handelsbanken is gratefully acknowledged. AI-based tools were used for language editing, spell-checking, and drafting support. All errors are my own.

# 1   Introduction

Corruption is a fundamental challenge to democratic governance and economic development (Mauro, 1995; Rose-Ackerman, 1997). A large literature examines how electoral accountability disciplines corrupt politicians (Besley, 2006; Persson and Tabellini, 2002), how voters respond to information about malfeasance (Ferraz and Finan, 2008; Chong et al., 2015), and whether institutions can curb rent-seeking behavior (Persson, Tabellini and Trebbi, 2003). Progress in these areas depends critically on reliable corruption measures, yet corruption is inherently hidden, making it difficult to observe and measure.

Traditional corruption indicators face various limitations including unpredictable sampling and reporting bias (Sequeira, 2012). Cross-national indices from Transparency International or the World Bank reflect aggregate perceptions and offer no subnational variation (Treisman, 2007). Local perception-based surveys, where available, may capture media attention or political mood rather than actual wrongdoing (Olken, 2009). As a result, recently researchers, among other tools to measure corruption, have relied on manually coded audit reports as a more objective basis for identifying corruption.

Brazil's random audit program has provided the foundation for influential work in this tradition (Ferraz and Finan, 2008, 2011; Litschig and Zamboni, 2018; Brollo et al., 2013; Timmons and Garfias, 2015), and similar hand-coded measures have been developed in Mexico (Larreguy, Marshall and Snyder, 2020) and Puerto Rico (Bobonis, Cámara Fuertes and Schwabe, 2016). However, manual coding is slow, costly, and difficult to scale, and even trained (human) coders exhibit inconsistency and judgment noise (Kahneman et al., 2016).

This paper proposes an automated measure that overcomes these constraints. Using the text of Brazil's audit reports for all municipalities audited between 2003 and 2011, I extract several variables linked to corruption including the number of irregularities pointed out by the auditors. I classify each irregularity as 'severe' or 'non-severe' using a dictionary based method which can also achieve strong performance (Jacobs and Rau, 1990; Hayes and Weinstein, 1990; Manning, Raghavan and Schütze, 2010). I also use other variables such as the number of lines, number of pages, and number of images included in the audit reports to create a corruption index using a principal component analysis (PCA). The first principal component retains 80% of the common variation and is the only component with an eigenvalue greater than one.

Following recommended practices in text-as-data validation (Grimmer and Stew-art, 2013; Gentzkow, Kelly and Taddy, 2019), I evaluate the measure using multiple benchmarks. First, I compare the automated index with independently produced hand-coded corruption counts by Ferraz and Finan (2011) and Timmons and Garfias (2015). In samples where human coders agree closely, the automated index aligns strongly with their judgments ($R^2 \approx 0.72$), indicating that it captures the same corruption-severity dimension that researchers identify. Second, I validate the index against the auditor's own classifications of severe irregularities, introduced in later audit rounds. The correlation with this measure is weaker ($R^2 \approx 0.31$) as auditors mainly classify severity based on potential fiscal impact. Taken together, these tests show that the index tracks corruption severity in a manner consistent with both human evaluation and official classifications.

As an additional construct-validity check, I examine how the index correlates with municipal characteristics that the corruption literature identifies as systematic risk factors –such as socioeconomic development (Rothstein and Holmberg, 2019; Serra, 2006), media penetration (Strömberg, 2004), demographic structure (Treisman, 2000), and geographic remoteness (Campante and Do, 2014). The correlations align with theoretical expectations yet explain only a small fraction of variation in the automated index (joint $R^2$ =0.17). This pattern indicates that the automated index captures meaningful information from the audit text that is not reducible to standard municipal characteristics, isolating a latent corruption dimension that is otherwise unobserved.

For robustness, I classify severe irregularities using supervised learning alternatives –logistic regression and Naive Bayes – trained on the available hand-coded labels. Replacing the dictionary-based classifier with either supervised approach yields nearly identical municipal rankings ($R^2 = 0.98$), indicating that all methods extract the same underlying corruption-relevant signal from the audit text. The dictionary–PCA approach is therefore consistent with supervised models but has the important advantage of not requiring labeled data. This makes it applicable in settings where high-quality training sets are scarce, noisy, or costly to produce.

The corruption index is scalable, reproducible, and cost-effective, enabling measurement across a much larger set of municipalities than previous hand-coded efforts. Unlike LLM-based approaches, the method does not depend on the stability of proprietary APIs, version updates, prompt specifications, which facilitates long-term replicability (Barrie, Palmer and Spirling, 2024; Argyle et al., 2025). The index has already been used by Muço (2025) to examine how information about mayoral corruption generates electoral spillovers across parties, jurisdictions, and electoral races, showing the practical value of the measure.

This paper is organized as follows: Section 2 details the audit program, Section 3 describes the construction of the automated corruption measure, Section 4 provides validation with hand-coded data as well as auditor's own classification, Section 5 provides robustness of the samples and method, and finally 6 concludes.

## 2 Audit Program

In April 2003, the Brazilian federal government initiated an anti-corruption program aimed at combating corruption in local governments. Implemented by the Controladoria Geral da Uniaõ (CGU), an independent federal agency, the program consists of randomly selecting Brazilian local governments for audit purposes through the national lottery machinery. Once a municipality is selected, a team of auditors located in each state in Brazil, visits the municipality to investigate the use of federal government transfers. After several days of on- and off-site investigation, the auditors compile a detailed report of their findings, which is made available for the public and the media on the CGU website. These reports include specifics about the inspection process, including back-and-forth with the mayors, as well as members of the municipal chamber, about the allegations and auditors frequently illustrate their findings using photographic evidence.

The audit program selected 50 - 60 municipalities for audit in each round, on average. Municipalities with a population above 500,000 were excluded from the lottery. From year 2003 - 2011, the years included in the sample, the auditors conducted 2194 inspections.[1] Starting from lottery round 8, the auditors included a detailed summary – of the irregularities found during inspections – in the first pages of the audit reports. The purpose of these summaries was media dissemination.

---

[1] Brazil has 26 states and Brasília where the Federal Capital is located. There are 5570 municipalities in Brazil.

# 3  Corruption Measure

## 3.1  Unsupervised Learning

Using the audit reports, and from lottery 8 the summaries provided by the auditors, I classify each irregularity as 'severe' or 'non-severe' – $c_h$ or $c_l$, respectively using a dictionary based method. Dictionary based approaches offer important advantages in transparency, interpretability, and reproducibility particularly when applied to documents in which expert knowledge can be translated into explicit and verifiable rules (Gentzkow, Kelly and Taddy, 2019).

A substantial body of work in information extraction shows that rule-based and dictionary methods can achieve strong performance (Jacobs and Rau, 1990; Hayes and Weinstein, 1990; Manning, Raghavan and Schütze, 2010). These settings include technical, legal, and administrative texts, where terminology is standardized and concept boundaries are well understood. Such conditions closely match the structure of CGU audit reports, making dictionary-based methods especially suitable for corruption measurement in this context.[2]

The dictionary was developed through iterative reading of a random set of audit reports across multiple lottery rounds with the help of a research assistant. After reviewing hundreds of reports to understand the range and language of irregularities, the dictionary consists of tokens and n-grams separate severe irregularities such as 'fraud' or 'ghost firm' from less severe such as 'missing (student) cadastral information'.[3] More details about the dictionary based method are discussed in the Appendix Section A2.

For each audit report, I count the number of irregularities, number of images, number of pages and the number of lines in each audit report, as well as number of severe irregularities using the dictionary based classification explained above. All these variables are right-skewed and exhibit high pairwise correlations. The pairwise correlations – among variables – range from 0.56 to 0.96 and are shown in Panel (a) of Figure 1. Each variable raw distribution is plotted in Appendix Figure A5.

To capture the underlying corruption in municipalities, I use a PCA. PCA is particularly appropriate in this analysis with highly correlated variables because not only han-dles multicollinearity naturally, but also requires no distributional assumptions about the underlying data. The first component accounts for approximately 80% of the variation in the data as shown in Panel (b) and is the sole component with an eigenvalue greater than one as shown in Panel (c).[4]

---

[2] A more detailed description of the procedure for extracting these summaries and the additional steps for harmonizing the information for earlier rounds of the lottery is provided in the Appendix.

[3] The dictionary is available at `https://github.com/ariedamuco/Audit-reports/blob/master/list-words`.

All variables load positively onto the first principal component, with loadings ranging from 0.37 to 0.48 (as shown in Appendix Table A1). Crucially, no single variable dominates the component, with the number of severe irregularities, report pages, number of images, lines, and total irregularities all contributing significantly. This balance confirms that the index is not merely a proxy for report length or any single metric, but rather a composite measure capturing the latent, underlying dimension of corruption severity in the audit reports.

The distribution of first principal component is shown in Panel (d) of Figure 1. The variable is right skewed and ranging from -2.93 to 14.34. Since the variables used for PCA are standardized to ensure comparability and no disproportional loading of any variable, the corruption measure is unitless. The lower end of the distribution indicates low corruption in municipalities, whereas the upper bound indicates high corruption.

# 4  Validation

I assess the validity of the corruption index in several complementary ways in line with Grimmer and Stewart (2013). First, I evaluate criterion validity by testing correspondence with independent human-coded corruption measures. Second, I examine external validity by showing that the index aligns with the official CGU classification of severe irregularities. Finally, I examine descriptive correlations between the index and municipal characteristics linked to corruption in prior research, showing that the measure behaves in theoretically consistent ways and is not reducible to municipality size or socioeconomic structure.

---

[4]Component selection is based on eigenvalue criteria and explained variance (James et al., 2013). Rules-of-thumb used to select the number of principal components for retention are: (i) retain components with eigenvalues larger than one, or (ii) ignore components with low additional variation. The variance explained by each eigenvalue and the cumulative variance are visualized.

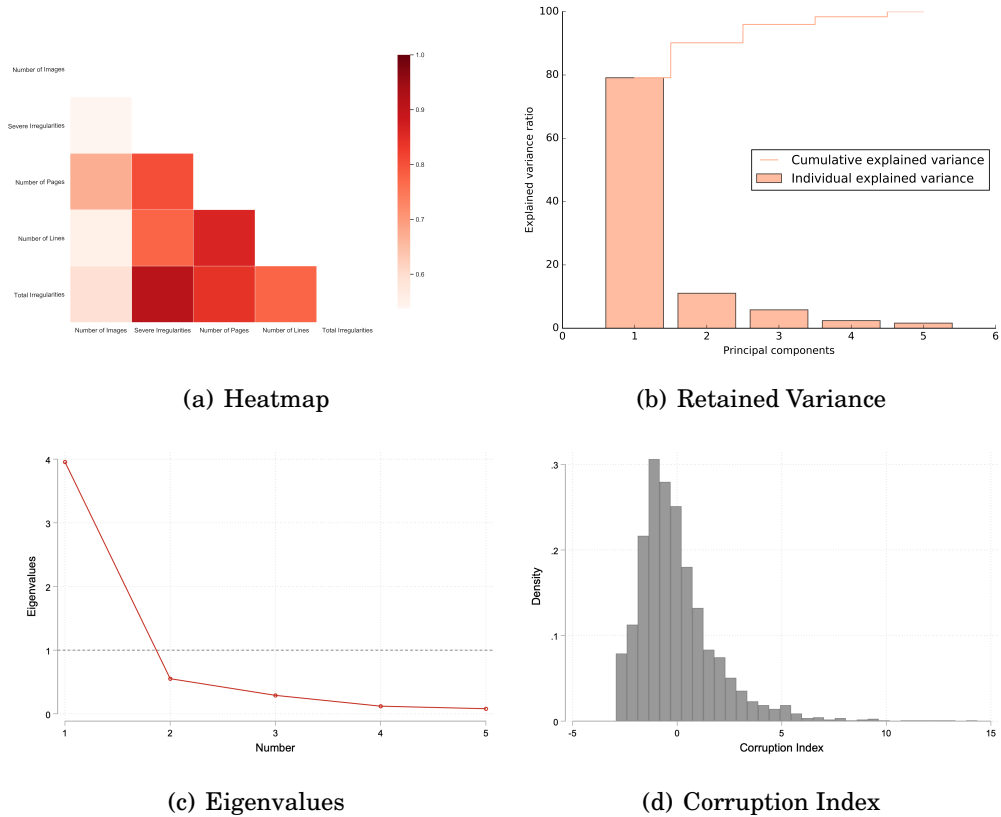

Figure 1: Principal Component Analysis

(a) Heatmap

(b) Retained Variance

(c) Eigenvalues

(d) Corruption Index

*Notes:* Panel (a) plots the raw pairwise correlation among the variables included in the PCA: the number of images; the number of severe irregularities, the number of pages, the number of lines, and the number of irregularities encountered by the auditors in the municipality. Panel (b) shows the individual explained variance of each component and the cumulative variance. The first principal component retains 80 percent of the explained variation. Panel (c) plots the number of components, in the x axis, and the respective eigenvalue in the y axis. Panel (d) plots the distribution of the corruption index (first principal component). The variable ranges between -2.93 and 14.34.

## 4.1 Validation with Human-Coded Data

For the first validation exercise, I use measures have been previously used in the literature by Ferraz and Finan (2011), henceforth FF, and Timmons and Garfias (2015), henceforth GT. The first are the pioneers of using this data. The 2011 measure is an improvement and expansion – of lottery rounds – of what was coded and used by the authors in their 2008 paper.[5] The second measure was coded and used by Timmons and Garfias (2015), attempting to approximate the coding in FF. Both variables count the instances of corrupt activities and are plotted in Appendix Figure A6.

I use a linear regression model to predict the hand-coded measures on the samples

---
[5]See namely Ferraz and Finan (2011).

that both coders agree, using automated corruption measure as follows:

$$HC_i = \alpha + \beta Corruption\ Index_i + \tau_t + \varepsilon_i \tag{1}$$

where $HC_i$ denotes hand-coded measure for municipality $i$, $Corruption\ Index_i$ is the PCA-based index constructed from audit-text irregularities and $\tau_t$ are state fixed effects to control for variation at the state level, such auditors team specific and language effect. For interpretability, the automated corruption measure is standardized.

I estimate variations of Equation 1. First, I focus on the subset of municipalities where both human coders are in complete agreement about the number of corruption irregularities. Second, I expand to the sample in which coders either fully agree or differ by at most one count. In both cases, the dependent variable is the mean of the two coders' scores.[6] Finally, I estimate the same specification each individual sample where the outcome variables are each coder's raw corruption count.

Table 1 present results. Column (1) reports the specification using the strict agreement sample, in which FF and GT assign the exact same number of corrupt activities to a municipality. Column (2) expands the sample to include municipalities in which coders differ by at most one count. Finally, Columns (3) and (4) report the same specification using the full sets of FF and GT coder labels, respectively.

Table 1: Predicting Human Coders

|  | FF = GT | |FF - GT| ≤ 1 | FF | GT |
| --- | --- | --- | --- | --- |
|  | (1) | (2) | (3) | (4) |
| Corruption Index | 1.492*** | 1.276*** | 0.628*** | 2.120*** |
|  | (0.432) | (0.197) | (0.108) | (0.202) |
| Observations | 34 | 99 | 462 | 364 |
| R2 | 0.726 | 0.714 | 0.382 | 0.488 |

Notes: The table reports coefficient estimates and, in parenthesis, heteroschedastic robust standard errors that result from a linear regression of the human coded measures, HC, on the automated corruption index conditioning for state fixed effects.
*** Significant at the 1 percent level.
** Significant at the 5 percent level.
* Significant at the 10 percent level.

---

[6] In the strict-agreement sample, this mean simply reproduces the common value assigned by both coders; in the broader sample, the mean provides a natural estimate of the underlying corruption level when assessments differ only marginally.

Columns (1) and (2) indicate that the automated measure has a strong predictive validity in the samples where human coders agree. A one–standard deviation increase in the corruption index is associated with 1.3–1.5 additional corrupt activities in the strict and near-agreement samples, with coefficient estimates tightly clustered and $R^2$ values between 0.71 and 0.73. This indicates that the index captures the underlying severity dimension where human assessments are most reliable.

When the full FF and GT datasets are used separately, namely columns (3) and (4), the coefficients remain positive and statistically significant - ranging from 0.6 to 2.1 - but the explanatory power declines (to 0.38 for FF and 0.49 for GT). Moreover, the corruption index predicts GT's corruption counts more closely than FF's, as reflected in the higher $R^2$. This pattern is consistent with the underlying hand-coded data. FF applies a more conservative coding rule, resulting in a compressed distribution, instead GT uses a wider scale. Thus, the corruption index explains more variation in the GT variable because the GT coder uses a broader scale.[7] These differences in predictive power reflect coder-specific scale choices, not necessarily differences in the predictive performance of the measure.

Overall, the pattern is consistent with the interpretation that the corruption index aligns most closely with human judgments where those judgments themselves are most reliable. However, the automated index cannot be expected to explain all variation in the human-coded measures even in the high-agreement samples. This is because, human measures are coarse, discrete counts recorded on a low-resolution scale (e.g., 0-10 or 0-19), whereas the automated index is continuous and incorporates substantially more information. Moreover, FF and GT may share common coding heuristics or biases, which raise inter-coder agreement but do not eliminate measurement error relative to the underlying corruption construct. As a result, the regression $R^2$ cannot approach 1, and the observed values (0.71 - 0.73) are consistent with a strong relationship once coder noise and scale limitations are accounted for.

## 4.2 Validation with CGU coded data

I validate the corruption against CGU's official classification of severe irregularities, available from lottery round 20 onward, which therefore does not overlap with the hand-coded samples from FF and GT. The CGU measure ranges from 5 to 95 (mean = 24.9, SD = 9.2) and records all irregularities classified as severe based on their potential financial losses to government. This contrasts with the academic hand-coded measures, which contain many zeros: the FF measure ranges from 0-10 (mean = 1.93, SD = 1.71) and the GT measure from 0-19 (mean = 3.76, SD = 3.12), each with substantial mass at zero.[8] As noted by Avis, Ferraz and Finan (2018), CGU classifies irregularities as severe based on potential fiscal impact, whereas academic coders weight acts by their nature and criminality.

---

[7] See Appendix Figure A6 for the distribution of each variable.

Table 2 reports the relationship between the corruption index and the CGU-coded measure. Column (1) uses the raw CGU count as the dependent variable, while Column (2) estimates the same specification using the logarithmic transformation of the CGU coded variable because the distribution of CGU coded irregularities is strongly right-skewed. The log specification down-weights extreme cases and captures proportional rather than absolute differences in corruption.

Because the dependent variable, the GCU corruption measure, exhibit a lognormal distribution with a long right tail, I test whether the results are robust to functional form by estimating specifications in both levels and logs. If outliers drive the relationship, results should differ substantially between these specifications. However, the finds suggest that the corruption index explains 29-31 % of the variation in CGU counts across both functional forms after conditioning on state fixed effects. The similarity of the results indicates that the index captures systematic variation in corruption severity rather than being driven by outliers.

Table 2: Predicting CGU

|  | CGU | log(CGU) |
| --- | --- | --- |
|  | (1) | (2) |
| Corruption Index | 4.584*** | 0.169*** |
|  | (0.527) | (0.014) |
| Observations | 926 | 926 |
| R2 | 0.311 | 0.293 |

Notes: The table reports coefficient estimates and, in parenthesis, heteroschedastic robust standard errors that result from a linear regression of the CGU coded measure on the automated corruption index conditioning for state fixed effects.
*** Significant at the 1 percent level.
** Significant at the 5 percent level.
* Significant at the 10 percent level.

---

[8] See Figure A6 in the Appendix for the full distribution of the raw variable.

The findings suggest that one standard deviation increase in the automated measure corresponds to 4.6 additional severe irregularities in CGU's classification (t = 8.70, p<0.001, $R^2$ = 0.31) explaining 31% of the variation ($R^2$ = 0.31). The lower $R^2$ compared to the human-coded results ($R^2 \approx 0.71 - 0.73$) reflects the differences in classification. Academic coders (FF and GT) weight corrupt acts by their nature and criminality, whereas the CGU classifies irregularities as severe based primarily on their potential fiscal impact. This distinction confirms the index aligns with different, yet officially relevant, dimensions of corruption severity.

## 4.3 Correlates of corruption and incremental validity

After showing the index aligns with human coders and CGU ratings, I next examine how it behaves relative to municipal characteristics commonly associated with corruption. This exercise serves as a construct-validity check rather than a determinants analysis, assessing whether the index displays theoretically consistent patterns without being reducible to municipality size or socioeconomic structure. Data on municipal characteristics are retrieved from the Brazilian Institute of Geography and Statistics (IBGE) and the Institute for Applied Economic Research (IPEA).

As an additional construct-validity exercise, I examine whether the corruption index displays the correlations that the literature consistently associates with corruption. Specifically, I consider socioeconomic development (literacy and GDP per capita), which prior work links to lower corruption (Rothstein and Holmberg, 2019); demographic and administrative structure (urbanization and population size), which shape opportunities for malfeasance (Treisman, 2000); local media penetration (FM radio availability), often associated with increased transparency and accountability (Strömberg, 2004); and geographic remoteness (distance to the federal capital), which has been shown to predict higher corruption in more isolated settings (Campante and Do, 2014).

Results are shown in Table 3. Across all bivariate specifications, shown in columns (1) - (6), the correlates behave in theoretically consistent ways. Literacy rates, GDP per capita, and urban population share are negatively associated with the automated corruption index, consistent with the idea that more developed and better-educated municipalities exhibit lower corruption risks. Population size (log) and distance to the Federal Capital are positively correlated with the index, indicating that larger municipalities and those farther from Brasília tend to receive higher corruption scores. FM radio presence is also positively correlated in the bivariate specification. While this is at first perhaps unexpected, it reflects the fact that municipalities with radio stations are typically larger, richer, and more administratively complex – features that increase the number of detectable irregularities – rather than a causal effect of media on corruption.

Table 3: Correlates of Corruption

|  | Corruption Correlates | | | | | | |
| --- | --- | --- | --- | --- | --- | --- | --- |
|  | (1) | (2) | (3) | (4) | (5) | (6) | (7) |
| Literacy rate | -0.050*** <br> (0.004) |  |  |  |  |  | -0.043*** <br> (0.004) |
| GDP per capita |  | -0.041*** <br> (0.012) |  |  |  |  | -0.010** <br> (0.004) |
| Urban population |  |  | -0.009*** <br> (0.002) |  |  |  | 0.001 <br> (0.002) |
| Population (log) |  |  |  | 0.301*** <br> (0.044) |  |  | 0.337*** <br> (0.046) |
| Distance to Federal capital |  |  |  |  | 1.146*** <br> (0.087) |  | 0.720*** <br> (0.088) |
| Radio FM |  |  |  |  |  | 0.293*** <br> (0.088) | 0.039 <br> (0.086) |
| Observations | 2136 | 2136 | 2136 | 2136 | 2136 | 2136 | 2136 |
| R2 | 0.104 | 0.016 | 0.012 | 0.027 | 0.075 | 0.006 | 0.172 |

Notes: The table reports coefficient estimates with heteroskedasticity-robust standard errors in parentheses. Columns (1)–(6) present bivariate regressions of the automated corruption index on each municipal characteristic separately. Column (7) reports the multivariate specification including all characteristics jointly. All variables are measured at the municipal level, reference year 2000. Literacy rate, GDP per capita, and urban population share are recorded directly by IBGE. Population (log) is the natural logarithm of total municipal population. Distance to the Federal Capital is the straight-line distance in kilometers between the municipality's population centroid and Brasília, scaled by 1/1000 for interpretability. Radio FM is an indicator equal to one if the municipality had an FM radio station in 2000.
*** Significant at the 1 percent level.
** Significant at the 5 percent level.
* Significant at the 10 percent level.

No single municipal characteristic explains more than about 10% of the variation in the corruption index, highlighting that the index is not reducible to standard socioeconomic or demographic factors. The $R^2$ ranges from 0.006 -0.1. Jointly the variables explain only 17% of variation in the index. In this multivariate specification, most coefficients retain the expected signs, although urban population share and FM radio become statistically insignificant.

Overall, these results show that while the automated index correlates with known correlates of corruption in theoretically consistent ways, it is by no means reducible to standard socioeconomic or demographic observables, indicating that the measure captures substantial independent information from the audit texts.

# 5 Robustness

## 5.1 Leave-one-out sensitivity analysis

Because the human-coded validation set is small, I implement a leave-one-out analysis in which Equation 1 is re-estimated after excluding one observation in each time. This allows me to assess whether individual audits exert undue influence on the results.

Figure 2 displays the leave-one-out re-estimations results for the two human-coded validation samples. The solid line plots the $R^2$ and the dots in gray plots the model's coefficient estimates. In Panel (a), full agreement or $FF = GT$, coefficient estimates range from 1.20 to 1.73, with $R^2$ values between 0.67 and 0.83. In Panel (b), $|FF - GT| \leq 1$, coefficient estimates range from 1.17 to 1.35, and $R^2$ ranges from 0.66 to 0.74. The limited variation indicates that the validation results are not driven by individual observations. the estimated coefficient and the model's explanatory power remain highly stable across iterations, indicating that no single audit observation drives the validation result.

Figure 2: Robustness: Leave one out

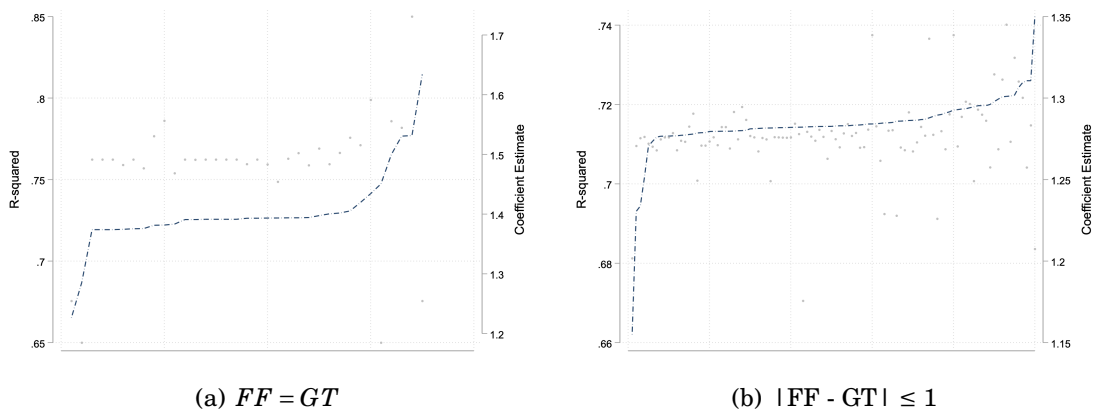

(a) $FF = GT$  (b) |FF - GT| ≤ 1

*Notes:* Each panel reports leave-one-out re-estimations of Equation 1. Gray dots show the estimated coefficient on the automated corruption index in each iteration, sorted from lowest to highest. The dashed line plots the corresponding $R^2$ using the same ordering. Coefficient estimates range from 1.17 to 1.35 in the $|FF - GT| \leq 1$ sample and from 1.20 to 1.73 in the $FF = GT$ sample, with $R^2$ ranging from 0.66 to 0.74 and 0.67 to 0.83, respectively. The limited variation indicates that no single audit observation drives the validation results.

## 5.2 Supervised Learning

I also use supervised learning approaches, specifically a logistic regression (LR) and Naive Bayes (NB), as alternative methods to obtain the severe irregularity count in each audit report. To train the algorithm, I use the human-coded measures as label data. I construct the target value to mimic high-versus low severity of corruption, hence I label as 'high corruption' for those municipality scoring at or above the sample median, and 'low corruption' for scores below the median. The labeled data come from the subset in which both independent coders, FF and GT, agree that the instance is above the sample median. The observations where the coders disagree are excluded.[9] The classification report for both classifiers is shown in Appendix Table A3.[10]

The trained classifier is then applied to identify text patterns associated with severe corruption, producing an alternative classification of irregularities that can be used in place of the dictionary-based counts.

Figure 3 plots the results from the LR classification. It shows the most common bigrams that contribute to each class. In gray, I show the top ten bigrams that are linked to low corruption. In orange, instead, I plot the top 10 bigrams linked to high corruption. *High corruption* bigrams are related to procurement processes and resource management, i.e., "procurement contract," "release resources," "proof expenditure," "fraud process," etc. On the other hand, *low corruption* bigrams are related to "divergence of information," "school lunch," or lack of certain formalities. These bigrams substantially overlap with the dictionary used for classification, suggesting that both the hand-crafted dictionary and the data-driven approach identify similar linguistic markers of corruption severity.

---

[9]To get labeled data for supervised learning classifications tasks, two independent are ask to perform the same classification task to create labeled data. The classification is accepted as accurate if the coders agree. If no agreement is reached, a third coder is asked to perform the task, and the procedure continues until an agreement is reached. In this instance I rely on two independent coders and due to resource constraints, I cannot involve a third coder.

[10]Since the purpose is to validate the text signal against human coders rather than to select a predictive model, performance metrics are computed on the full hand-coded set. Results are similar under a 70/30 split.

Figure 3: Top bigrams for each class

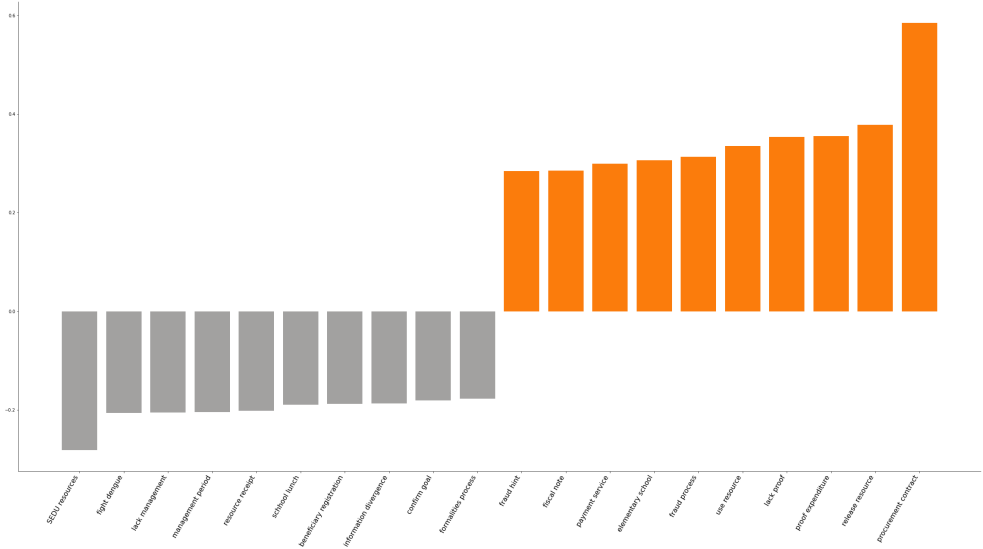

*Notes:* Top 10 bigrams that contribute to each class, high or low corruption. An instance is classified as high corruption if it scores above or equal to the sample median in the human-coded sample. An instance that scores below the sample median is classified as low corruption. Most frequent bigrams in low corruption reports are represented in gray. In orange, I plot bigrams linked to high corruption.

I also construct an alternative PCA index replacing dictionary-derived severe irregularity counts with predictions from machine learning classifiers. The correlation between dictionary-PCA and LR-PCA is above 0.98, indicating that both methods produce nearly identical municipal corruption rankings despite using different classification approaches.[11] This near-perfect convergence provides is likely due to the use of the PCA after the text classification and provides strong evidence that the measure captures robust variation in corruption severity.

Supervised learning offers a natural alternative if independent training set is available. In principle, corruption classifications could be learned directly from hand-coded audit reports. However, this approach depends on large and reliable training sets, while the available labels are both limited and inconsistent. As a result, supervised models capture broad differences in severity but remain sensitive to coder disagreement and cannot be extended to the full set of audits.

---

[11] The high convergence is consistently observed regardless of whether a NB or LR classifier is used to generate the predicted severe irregularity counts.

The fact that a PCA constructed from supervised-derived features is highly correlated with the dictionary PCA index is reassuring and shows that when labels exist and are consistent, both approaches recover the same underlying corruption dimension. This strong consistency across fundamentally different classification approaches strengthens the conviction that the dictionary-PCA method successfully isolates the robust underlying corruption dimension present in the audit text.

# 6 Conclusion

This paper introduces a transparent and reproducible method for measuring corruption using the full text of Brazil's municipal audit reports. The approach classifies severe irregularities with a domain-specific dictionary and aggregates multiple indicators into a single corruption-severity index via principal component analysis. Supervised-learning alternatives produce nearly identical results, underscoring that the dictionary–PCA method captures the same latent dimension while retaining interpretability and replicability.

Validation exercises demonstrate that the automated index closely aligns with independent expert coders, explaining 71–73% of the variation in human-coded corruption where coder agreement is highest. The index also predicts the CGU's administrative classifications of severe irregularities, accounting for roughly 30% of their variation. A third validation shows that the index correlates with municipal characteristics consistently linked to corruption risk, though these factors explain only a modest share of its variation – indicating that the measure captures information not reducible to standard socioeconomic or geographic observables.

The index has limitations: it captures only irregularities documented by auditors, may partially reflect administrative capacity, and depends on the scope and thoroughness of the audit process. These constraints, however, apply equally to hand-coded datasets and LLM-based approaches. Within these boundaries, the automated method reduces subjective judgment, minimizes coding noise, and provides a consistent, text-anchored measure of corruption severity.

Because the methodology is scalable, validated, and fully reproducible, expanding available samples by a factor of three to five relative to hand-coded datasets. This broader coverage has already enabled fine-grained analyses of how corruption information affects political behavior across jurisdictions, and it creates opportunities for new work on account-ability, governance, and institutional performance. The method can be readily extended to other audit-based settings, providing a foundation for future research on corruption, governance, and institutional performance.

# Appendix for Measuring Corruption from Text Data

Arieda Muço[*]

***Appendix***

# A1 Details about the audit program

## A1.1 Report summaries

To facilitate media dissemination, from lottery 8, the CGU decided to include summaries of the findings in the first pages of the audit reports. The auditors enlist the irregularities found for each federally transferred fund and the respective ministry in charge in these summaries. An extract of a summary is provided in Figure A1 from the municipality Curiuva, state of Paraná, audited in lottery round 11.

---

[*]Central European University. E-mail: arieda.muco@gmail.com

## Figure A1: Summary Example

Ministério da Agricultura, Pecuária e Abastecimento:

1.1. Inobservância de preceito da Lei nº 9.452/97 pela Prefeitura Municipal;
1.2. Desorganização formal de documentação referente a processo licitatório, em descumprimento ao disposto no artigo 38 da Lei nº 8.666/93;
1.3. Realização de licitação na modalidade Convite quando caberia Tomada de Preços; e, simulação de realização de processo licitatório;
1.4. Guarda inadequada de trator adquirido pelo Programa PRODESA.

Ministério das Cidades:

1.1. Inobservância às exigências do artigo 38 da Lei nº 8.666/93 quanto à formalização de processo;
1.2. Inobservância de preceito da Lei nº 9.452/97 pela Prefeitura Municipal.

Ministério das Comunicações:

1.1. Inexistência de postos de atendimento pessoal a usuários do serviço telefônico fixo comutado – STFC;
2.1. Restrição do acesso à Internet.

Notes: Example of a summary in the municipality Curiuva in the state of Paraná in lottery round 11. For each ministry, the auditors enlist the irregularities found. Lei nº 9.452/97 and Lei nº 8.666/93 indicate the law on public procurement and announcement of reception of federal funds, respectively.

## A1.2 Example of a Summary (Translated)

Ministry of Agriculture, Livestock, and Supply:

1.1 Non-compliance with the precept of Law n.9.452/97 by the City Hall.

1.2 Formal disorder of documents referring to the bidding process, in non-compliance with the dispositions of article 38 of the Law n. 8.666/93.

1.3 Execution of bidding in the modality 'Invitation' (Convite) when it should have been 'Submission of prices' (Tomada de preços) and, simultaneously, simulation of the execution of a bidding process.

1.4 Improper safekeeping of a tractor obtained by the program PRODESA.

Ministry of Cities:

1.1 Non-compliance to the requirements of article 38 of the Law n. 8.666/93 in relation to the formalization of a process.

1.2 Non-compliance with the precept of Law n.9.452/97 by the City Hall.

Ministry of Communications:

1.1 Non-existence of personal service stations to users of Fixed Switched Telephone Services -STFC (Serviço Telefônico Fixo Comutado).

2.1 Restriction on access to the Internet.

## A1.3   Identifying the relevant paragraph in early lotteries

Report summaries are not included up until lottery 8. The bulk of these reports, however, include summaries in the main text. To access this information, I try to locate the paragraph where the irregularity is described. In most reports, the summary of the findings is included before the description of the irregularity – it can be found between "Constatação da Fiscalização" (Inspection Findings) and "Fato" (Fact) or "Evidencia" (Evidence).

Panel (a) of Figure A2 shows a description of the irregularity, in bold letters, and it is located between the "Constatação da Fiscalização" and "Fato" keywords. I use regular expressions to extract the text between these keywords. Panel (b), instead, depicts an example of instances in which the auditors do not summarize the irregularity. For these instances, I construct the irregularity count by locating paragraphs that start with "Constatação da Fiscalização" or "Fato" and use the first sentences for each paragraph.

Figure A2: Differences in audit reports

(a) Irregularity with a summary.

1.1) Constatação da Fiscalização:

**Deficiência no pagamento automático da Bolsa Criança Cidadã, realizado via cartão magnético.**

**Fato**

Encontrou-se o programa em questão em bom andamento no Município no que tange a sua gestão. O Conselho Municipal de Assistência Social tem uma participação ativa tornando, desta forma, as ações sociais de fácil acesso à comunidade. Quanto à Comissão Municipal de Erradicação do Trabalho Infantil – CMETI, esta foi regularmente criada e encontra-se em funcionamento.

(b) Example of an irregularity without a summary.

1.1) Constatação da Fiscalização:
**Fato:**
Os veículos utilizados pelo Município nas atividades de endemias são insuficientes em número para atender as necessidades do programa e não são adequados pelos seguintes fatos: são veículos de capacidade reduzida para conduzir toda a equipe, para fazer o transporte do material a ser utilizado e por não chegarem a áreas de difícil acesso pelo fato de não ter tração.

**Evidência:**
O fato foi evidenciado por meio de inspeção "in loco" dos veículos e entrevista junto ao pessoal envolvido no programa.

Notes: Panel (a) depicts an irregularity with a summary. The description is in bold letters and located between the "Constatação da Fiscalização" and "Fato" keywords. Panel (b) depicts an irregularity without a summary where the auditors only describe their findings.

## A1.4 Fraud in Procurement

The Brazilian legislation – law number 8.666/93 – requires submitting at least three bids from different suppliers for a valid tender procedure. Below, are provided two instances of fraud in procurement processes where law requirements were circumvented.

The first example is from the municipality of Alvaraes, state of Amazonas, audited in round 2 of the lottery. The auditors documented four fraudulent procurement cases. In one instance, the call for bid was for cleaning products, but all three submitted bids were about food supply. The bids submitted have the same layout and mistakes. However, the auditors point out two main differences among the bid proposals: the first is the use of different fonts, and the second is related to the prices included in the bid. The photographic evidence included in the audit report is shown in Figure A3.

The second example is from the municipality of Afonso Cunha in Maranhão, audited in round 27. During the inspection, the auditors detected three fraudulent processes. In one tender, three firms were recorded to have submitted their proposals, but the auditors could not physically locate any of the firms including the company that was declared winner (CNPJ 07.981.189/0001-90). Locals also reported not knowing of the company. Additionally, the first installment for the service was paid by the municipality before the formalization of the bidding. Photographic evidence, as provided by the auditors, is shown in Figure A4.

Figure A3: Fraud in Procurement (1)

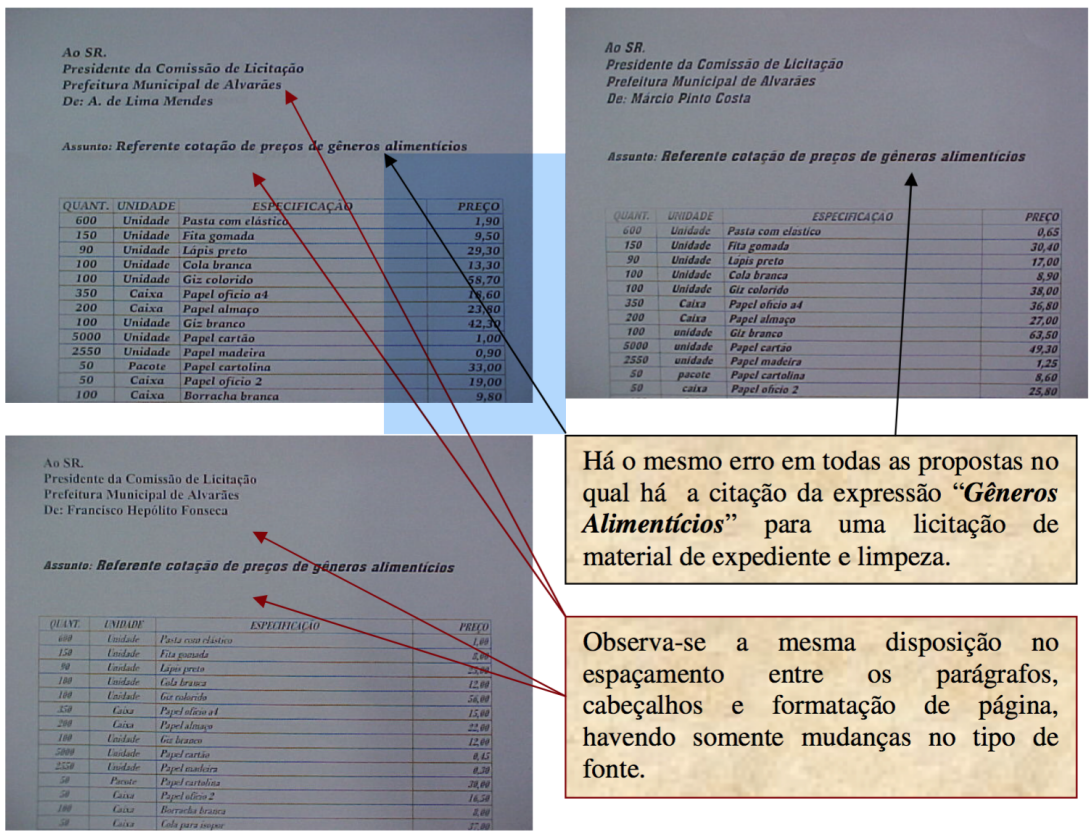

Notes: The call for bids was about cleaning products, but the proposals submitted are about food supply. The main differences across the bids are the font use of fonts and inflated prices.

Figure A4: Fraud in Procurement (2)

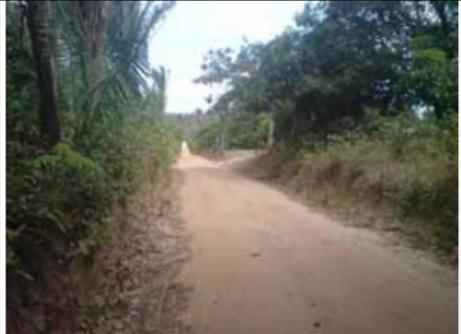
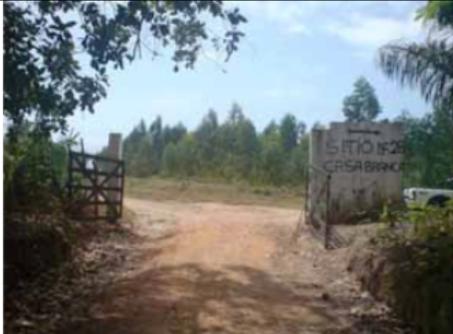
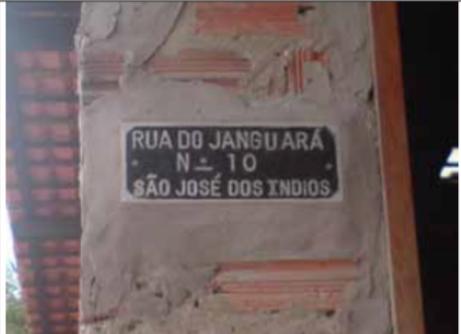
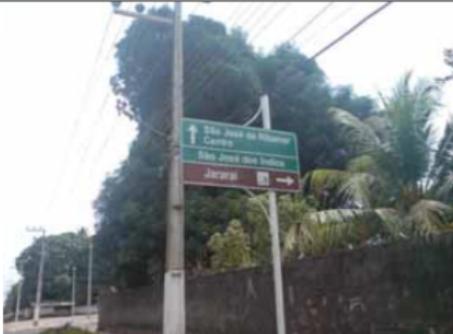
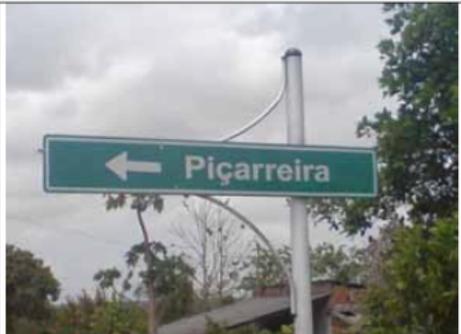
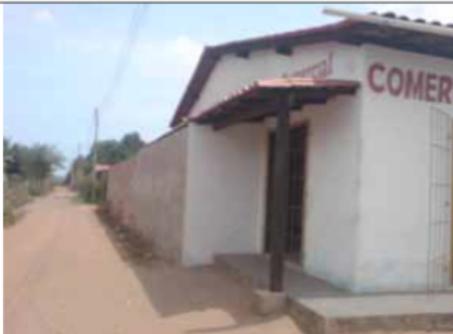

Notes: The auditors could not locate the firms indicated in the tender proposal. Under Brazilian law, all operating firms must maintain a registered business name and address and hold a unique taxpayer identification number. This identifier, is the national registry of legal entities in Brazil, Cadastro Nacional da Pessoa Jurídica (CNPJ), and is required for tax, regulatory, and commercial purposes.

# A2 Dictionary-based method

I construct a dictionary containing both single tokens and $n$-grams.[1] The raw text is preprocessed by removing accents, converting to lower case, and applying stemming, such that all tokens are analyzed in root form. To account for cases where different surface forms refer to the same underlying irregularity, the dictionary includes common synonyms and regular expressions.

Core corruption terms such as "fraud," "collusion," and "fake" are captured through single-token stems (e.g., `fraud`, `conluio`, `fals`). In addition, I include stemmed bigrams or trigrams to capture contextual expressions. For example, the English concepts "procurement simulations," "ghost firms," and "goods non located" map to the following Portuguese $n$-grams in the dictionary:

| English phrase | Portuguese equivalents in the dictionary |
|---|---|
| procurement simulations | *simulacao licitat*; *montagem licitat*; *licitat montad*; *processos pagamento simulado* |
| ghost firms | *empresa fantasma*; *empresa inexistente*; *empresa nao localizad*; *firm nao localizad* |
| goods non located | *ben nao localizad*; *bem nao localizad*; *desaparecimento materiais* |

*Notes:* This table provides examples of how English corruption-related expressions correspond to stemmed Portuguese $n$-grams used in the dictionary.

Non-located or non-existing firm, including ghost firms are all instances related to fraudulent procurement processes. A more detailed discussion on procurement simulations can be found in Section A1.4. In each audit report, I search if any of the n-grams appears in the description of the irregularity.[2] The instance is classified as severe irregularity if at least one n-gram is found. Finally, for each report, I count the total number of irregularities as well as severe irregularities.

---

[1] The dictionary was compiled with the assistance of a Brazilian research assistant after reading hundreds of audit reports; the complete list is available at https://github.com/ariedamuco/Audit-reports/blob/master/list-words.

As an illustration, consider the audit summary for municipality $X$ is the following list of identified irregularities. The classification generated by the algorithm is shown in square brackets:

1. Indication of fraud in the procurement process.  [Severe Irregularity]

2. Resources not used as designated by the program.  [Severe Irregularity]

3. Payments for unexecuted services.  [Severe Irregularity]

4. Non-actualization of pupils' cadastral information.  [Non-severe Irregularity]

5. Delay in the delivery of books to a rural school in the municipality.  [Non-severe Irregularity]

In this example, the total number of irregularities encountered by the auditors is five, and the number of severe irregularities is three.

## A2.1  Distribution of the variables

In Figure A5, I show the variables that I generate from the text of audit reports, such as the number of pages, the number of lines, the number of images (photographic evidence included in the audit report), and the number of total irregularities (the count of all irregularities summarized by the auditors for each municipality). All the variables have a long right tail and are highly correlated.

---

[2]A regular expression is a sequence of characters that define a search pattern, mainly for use in pattern matching with strings.

Figure A5: Histogram Plots

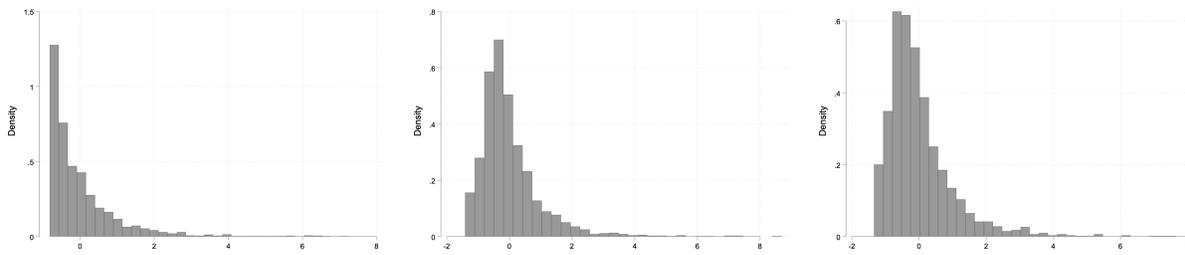

(a) Number Images  (b) Number Pages  (c) Number Lines

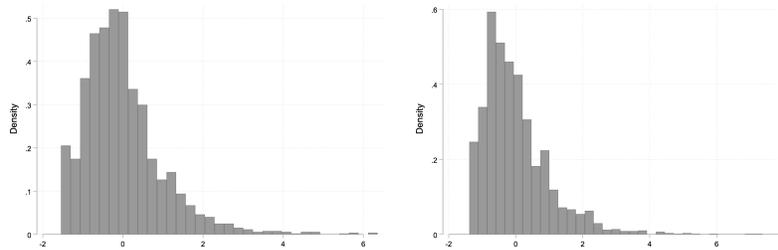

(d) Number of Irregularities  (e) Number Severe Irregularities

*Notes:* The figure plots a histogram for each standardized variable included in the algorithm: the number of images, the number of pages, the number of lines, and the number of irregularities encountered by the auditors in the municipality as well as the number of severe irregularities (resulting from text classification).

## A2.2 PCA-Loadings

Table displays the loadings of each principal component. Each column (1) - (5) shows the first through fifth principal components, namely PC1 - PC5, respectively. Loadings on PC1 are positive and similar in magnitude (0.37 - 0.47), indicating that this component reflects a common corruption-severity dimension rather than dominance by a single variable

Table A1: PCA Loadings

|  | PC1 | PC2 | PC3 | PC4 | PC5 |
|---|---|---|---|---|---|
| Number of Images | .371 | .900 | .129 | .189 | .008 |
| Severe Irregularities | .460 | -.325 | .460 | .356 | -.587 |
| Number of Pages | .475 | .010 | -.323 | -.745 | -.338 |
| Number of Lines | .452 | -.191 | -.690 | .487 | .215 |
| Total Irregularities | .470 | -.218 | .438 | -.212 | .704 |

Notes: Notes: PC1 - PC5 denote the first through fifth principal components.

# A3 Hand-Coded Measures

Figure A6 plots the raw distribution of corruption counts from FF, TG and CGU. Panel (a) shows the full FF distribution (N=476). Panel (b) instead shows the full GT distribution (N=369). Panel (c) shows the full CGU distribution (N=934).

Figure A6: Raw Hand-Coded Measures

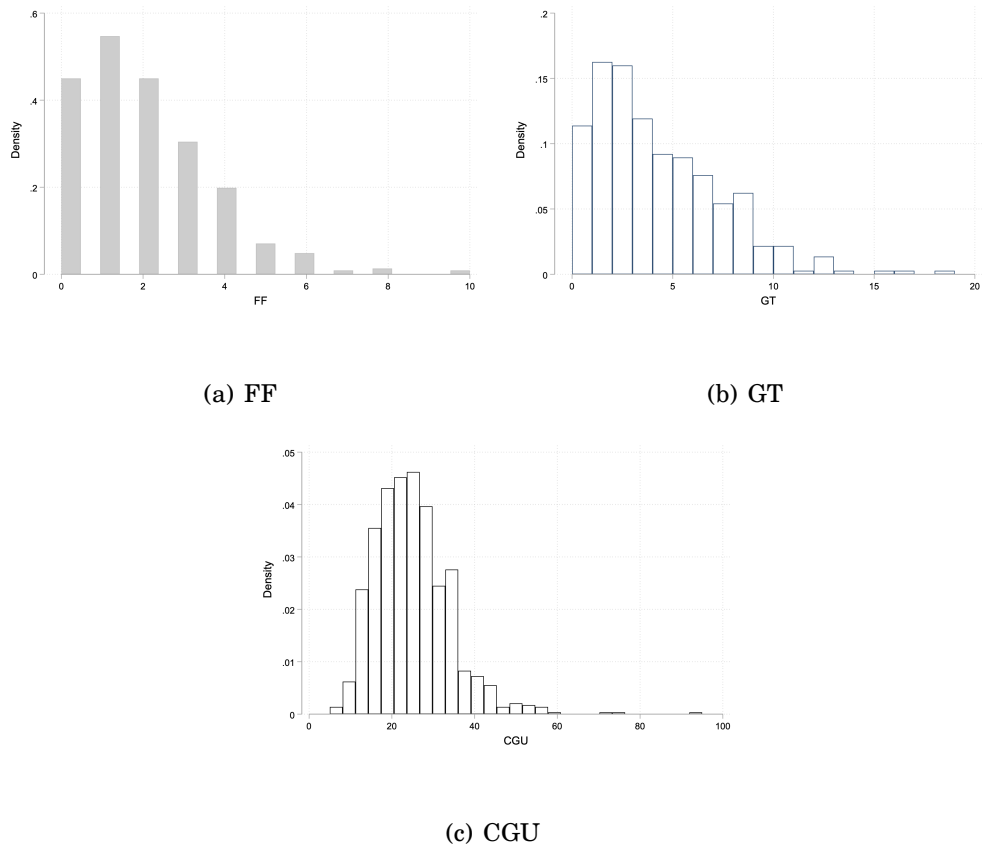

(a) FF

(b) GT

(c) CGU

*Notes:* Panel (a), Panel (b) and Panel (c) shows the full FF, GT and CGU distribution, accordingly.

Figure A7 plots the distribution of the outcome variable for validation in Section 4. Panel (a) plots the distribution of the hand-coded measure where coders assigned identical scores (N=34). Panel (b) instead shows the distribution of the average corruption count where coders agree within one unit (N=99). The overlap with the CGU distribution is omit-ted as it is identical to the raw variable plotted in Panel (c) of FigureA6.

Figure A7: Hand-Coded Measures: Outcome Variables

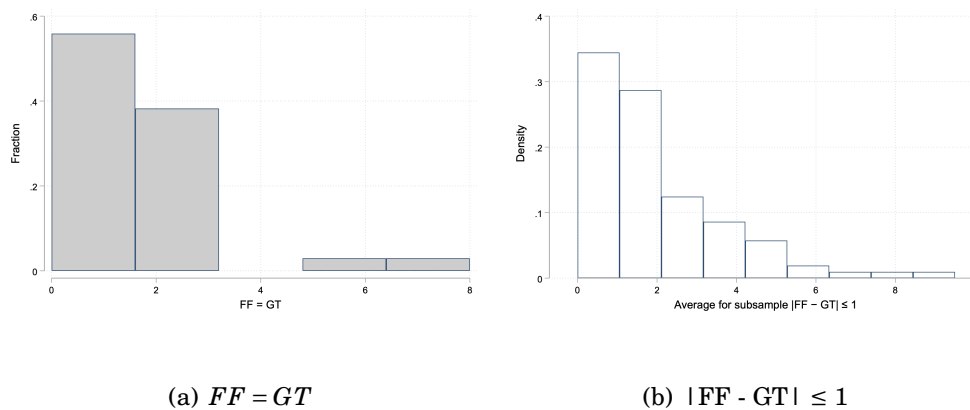

(a) $FF = GT$  (b) $|FF - GT| \leq 1$

*Notes:* Panel (a) shows the distribution of the hand-coded measure where coders assigned identical scores. Panel (b) shows the distribution of the average corruption count in the sample where coders agree within one unit.

The pattern of coder agreement reveals that independent experts reach consensus most reliably at the extremes of the distribution: 29% agree on zero corruption, 56% agree on 1-2 instances, and 6% agree on extreme corruption (6+ instances). The middle range (3-5 instances) accounts for only 9% of perfect agreement cases, suggesting this range involves greater subjective judgment.

This pattern has implications for corruption measurement more generally. Cases at the extremes (very low or very high corruption) appear to have clearer signals that expert coders identify consistently. Cases in the moderate corruption range show less consistency even among expert coders, suggesting this range is inherently more ambiguous. The automated measure's strong performance ($R^2$ = 0.70 - 0.73) despite this ambiguity indicates it captures the underlying corruption signal that human coders detect.

Table A2: Predicting Human Coders

|  | |FF - GT| ≤ 1 | | CGU | |
| --- | --- | --- | --- | --- |
|  | (1) | (2) | (3) | (4) |
| Corruption Index |  | 1.261*** |  | 4.578*** |
|  |  | (0.218) |  | (0.526) |
| Radio FM | -0.330 | -0.320 | 0.799 | 0.912 |
|  | (0.334) | (0.251) | (0.684) | (0.631) |
| Literacy rate | 0.013 | 0.019 | -0.052 | -0.011 |
|  | (0.024) | (0.020) | (0.056) | (0.051) |
| Urban population | 0.013 | -0.001 | -0.007 | -0.014 |
|  | (0.009) | (0.007) | (0.015) | (0.014) |
| GDP per capita | -0.035** | 0.004 | -0.182*** | -0.128*** |
|  | (0.017) | (0.012) | (0.068) | (0.048) |
| Distance to Federal capital | 2.011 | 1.173 | -1.997 | -1.233 |
|  | (1.207) | (0.954) | (1.665) | (1.629) |
| Population (log) | 0.271 | -0.020 | 0.603 | -0.130 |
|  | (0.219) | (0.165) | (0.423) | (0.333) |
| Observations | 99 | 99 | 920 | 920 |
| R2 | 0.555 | 0.729 | 0.173 | 0.318 |

Notes: The table reports coefficient estimates and, in parenthesis, heteroschedastic robust standard errors that result from a linear regression of the human coded measures on the automated corruption index conditioning for state fixed effects.
*** Significant at the 1 percent level.
** Significant at the 5 percent level.
* Significant at the 10 percent level.

# A4   Supervised Learning Approach

To apply machine learning algorithms to text data, I perform the necessary preprocessing steps. First, I remove words that add little meaning to the text. In information retrieval jargon, stopwords are the most common words in a language. (In English these would be `of`, `to`, `on`, `the`, etc.) I use the Portuguese stopwords provided by the Natural Language Toolkit (NLTK), a leading platform for building Python programs to work with human language data.[3]

Next, I stem the tokens using Porter Stemmer from the NLTK library to map each token to a common root ensuring equivalence among stems. For example, tokens "licitatorios" and "licitatorio" are mapped to "licitatori." Results are robust to performing other types of stemming such as Snowball or Lancaster stemmer and lemmatization. Finally, I consider bigrams – pairs of words – and assign weights using the term-frequency inverse document frequency (tf-idf) approach. The formula for the tf-idf is:

$$\text{tf-idf}_{x,y} = tf_{x,y} \times \log(\frac{N}{df_x}) \tag{1}$$

where $tf_{x,y}$ is the frequency of term $x$ in document y, $df_x$ is the number of documents containing $x$. Finally, $N$ is the total number of documents. For implementation, I use the TfidfTransformer from the scikit-learn library in Python (Pedregosa, Varoquaux, Gramfort, Michel, Thirion, Grisel, Blondel, Prettenhofer, Weiss, Dubourg, Vanderplas, Passos, Cournapeau, Brucher, Perrot, and Duchesnay, 2011).

I use Support Vector Machines with linear kernel and Naive Bayes algorithms to classify each instance of irregularity (Schütze, Manning, and Raghavan, 2008). Both algorithms exhibited robust performance, correctly classifying 96% of cases, a result consistent in both in-sample and out-of-sample evaluations. Results are shown in Table A3. Both display similar performance: have high precision, recall, and f1-score. Precision – positive predictive value – is the fraction of relevant instances among the retrieved instances. Recall – or sensitivity – is the fraction of relevant instances that were retrieved. F1-score is a weighted average of precision and recall. All these measures range between zero and one: the higher the score, the better the prediction. As a rule of thumb, in the machine learning literature, satisfactory metrics for these measures are above 0.8.

---

[3]For more information on the NLTK library and Natural Language Processing with Python. See Bird, Klein, and Loper (2009).

As shown in Table A3 all the metrics indicate a good performance of the algorithm. As expected, the algorithms are overfitting due to the small number of documents. This is expected due to the high dimensionality of text data and the small training sample. Hence, the bigrams resulting from the supervised learning method overlap with the dictionary based method. To get the equivalent corruption measure using this text classification, I proceed and count the severe irregularities emerging from the classification and use the variables in a PCA. The raw correlation between the corruption measures, the first resulting from the dictionary classifier and the second from the classifiers using supervised learning is high, above 0.98.

Table A3: Classification Report

Panel A: Naive Bayes

|  | Precision | Recall | F1-Score | Support |
|---|---|---|---|---|
| Low Corruption | 1.0 | 1.0 | 1.0 | 50 |
| High Corruption | 1.0 | 1.0 | 1.0 | 83 |
| Accuracy |  |  | 1.0 | 133 |
| Macro Avg | 1.0 | 1.0 | 1.0 | 133 |
| Weighted Avg | 1.0 | 1.0 | 1.0 | 133 |

Panel B: SVM (linear kernel)

|  | Precision | Recall | F1-Score | Support |
|---|---|---|---|---|
| Low Corruption | 1.0 | 0.84 | 0.91 | 50 |
| High Corruption | 0.91 | 1.0 | 0.95 | 83 |
| Accuracy |  |  | 0.94 | 133 |
| Macro Avg | 0.96 | 0.92 | 0.93 | 133 |
| Weighted Avg | 0.95 | 0.94 | 0.94 | 133 |

Notes: Precision is the fraction of relevant instances among the retrieved instances. Recall is the fraction of relevant instances that were retrieved. F1-score is a weighted average of precision and recall. Support is the number of instances that support the class.

## A4.1 Individual samples: Results

The performance of the algorithms in each individual sample, FF and GT, is shown in Table A4 and Table A5, respectively. High and low corruption are assigned with respect to the sample median.

In Figure A8, I plot the 20 most common features – bigrams – that contribute to each class for each separate sample. Gray and orange represent low and high corruption, respectively. Top contributing bigrams are related to the procurement processes and resource management, i.e. "procurement contract," "release of resource," "lack of expenditure proof," "fraud procurement process.". Low corruption instances are related to the divergence of information and school lunches or lack of certain formalities. Bigrams are stemmed and included in the original language.

Results are comparable across algorithms and samples.

Table A4: Classification Report (FF sample)

Panel A: Naive Bayes

|  | Precision | Recall | F1-Score | Support |
|---|---|---|---|---|
| Low Corruption | 0.96 | 0.99 | 0.98 | 221 |
| High Corruption | 0.99 | 0.97 | 0.98 | 245 |
| Accuracy |  |  | 0.98 | 466 |
| Macro Avg | 0.98 | 0.98 | 0.98 | 466 |
| Weighted Avg | 0.98 | 0.98 | 0.98 | 466 |

Panel B: SVM (linear kernel)

|  | Precision | Recall | F1-Score | Support |
|---|---|---|---|---|
| Low Corruption | 0.93 | 1.0 | 0.96 | 221 |
| High Corruption | 1.0 | 0.93 | 0.96 | 245 |
| Accuracy |  |  | 0.96 | 466 |
| Macro Avg | 0.96 | 0.96 | 0.96 | 466 |
| Weighted Avg | 0.96 | 0.96 | 0.96 | 466 |

Notes: Classification reports, Naive Bayes and Support Vector Machines, using Ferraz and Finan (2011) sample.

Table A5: Classification Report (GT sample)

Panel A: Naive Bayes

|  | Precision | Recall | F1-Score | Support |
|---|---|---|---|---|
| Low Corruption | 1.0 | 0.91 | 0.95 | 156 |
| High Corruption | 0.94 | 1.0 | 0.97 | 203 |
| Accuracy |  |  | 0.96 | 359 |
| Macro Avg | 0.97 | 0.96 | 0.96 | 359 |
| Weighted Avg | 0.96 | 0.96 | 0.96 | 359 |

Panel B: SVM (linear kernel)

|  | Precision | Recall | F1-Score | Support |
|---|---|---|---|---|
| Low Corruption | 1.0 | 0.96 | 0.98 | 156 |
| High Corruption | 0.97 | 1.0 | 0.98 | 203 |
| Accuracy |  |  | 0.98 | 359 |
| Macro Avg | 0.98 | 0.98 | 0.98 | 359 |
| Weighted Avg | 0.98 | 0.98 | 0.98 | 359 |

Notes: Notes: Classification reports, Naive Bayes and Support Vector Machines, using Timmons and Garfias (2015) sample.

Figure A8: Top-bigrams by class

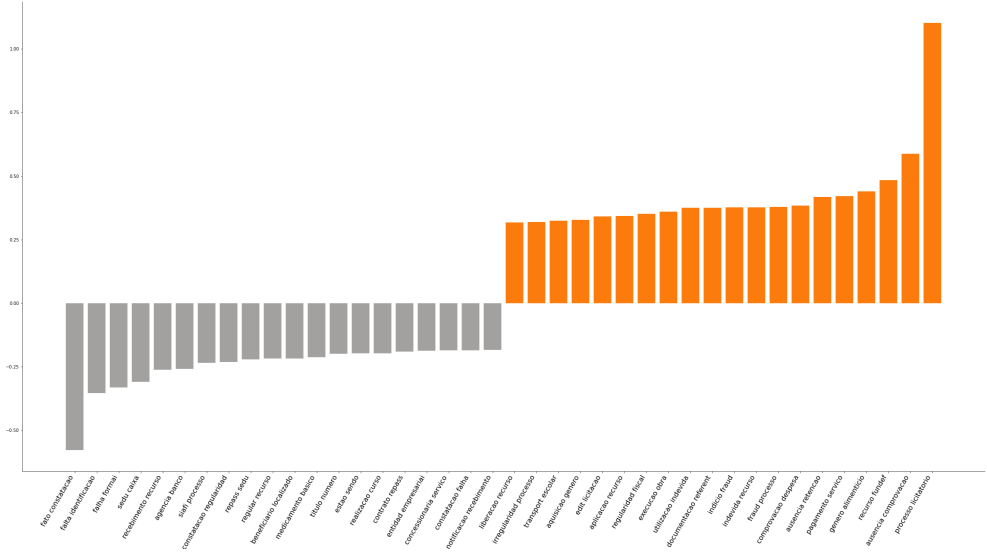

(a) FF sample

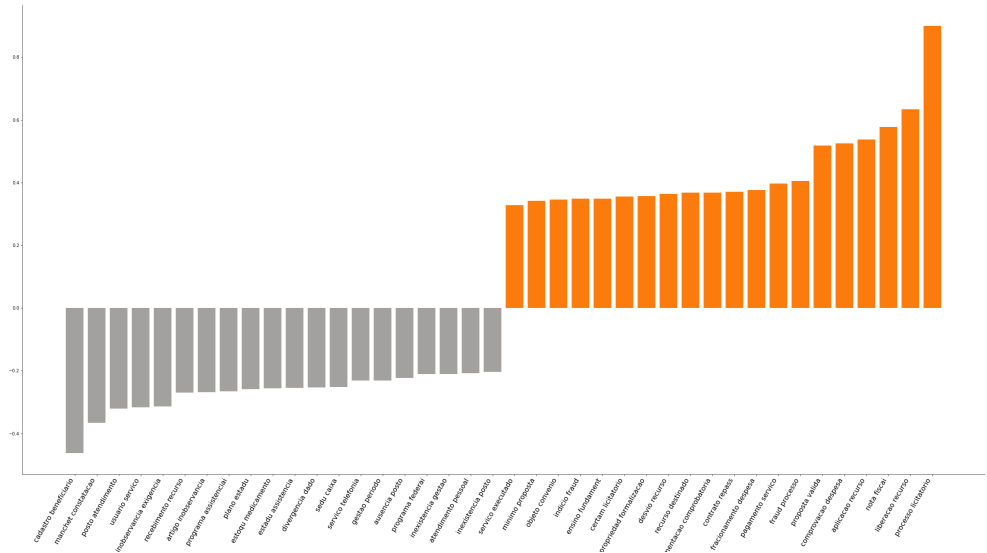

(b) GT sample

Notes: Panel (a) and (b) depict top 20 bigrams that contribute to each class – high or low corruption – shown in gray or orange, respectively.